\documentclass{emulateapj}
\usepackage{amsmath,natbib,psfig}
\input{buote.defs}

\def\stacksymbols #1#2#3#4{\def\theguybelow{#2}
        \def\verticalposition{\lower#3pt}
        \def\spacingwithinsymbol{\baselineskip0pt\lineskip#4pt}
        \mathrel{\mathpalette\intermediary#1}}
\def\intermediary #1#2{\verticalposition\vbox{\spacingwithinsymbol
        \everycr={}\tabskip0pt
        \halign{$\mathsurround0pt#1\hfil##\hfil$\crcr#2\crcr
                \theguybelow\crcr}}}

\begin{document} 

\title{Ultra-Low Iron Abundances in Distant Hot Gas in Galaxy 
Groups}
\author{David A. Buote\altaffilmark{1}, Fabrizio
Brighenti\altaffilmark{2,3}, \& William G. Mathews\altaffilmark{2}}
\altaffiltext{1}{Department of Physics and Astronomy, University of
California at Irvine, 4129 Frederick Reines Hall, Irvine, CA 92697-4575}
\altaffiltext{2}{UCO/Lick Observatory, Department of Astronomy
and Astrophysics, University of California, Santa Cruz, CA 95064}
\altaffiltext{3}{Dipartimento di Astronomia, Universit\`a di Bologna,
via Ranzani 1, Bologna 40127, Italy}

\slugcomment{To Appear in The Astrophysical Journal Letters}

\begin{abstract}
A new \xmm\ observation of the outer regions of the galaxy group NGC
5044 indicates hot gas iron abundances of only $\fe/\solar \sim 0.15$
between $r=0.2 - 0.4\, \rvir$.  While the total baryon mass within the
virial radius may be close to the cosmic mean value observed in rich
clusters, the ratio of total iron mass to optical light in NGC 5044 is
about 3 times lower than that in rich clusters.  The remarkably low
iron abundance over a large volume of the intergroup gas in the outer
regions of NGC 5044 cannot be easily understood in terms of the
outflow of enriched gas in a group wind during its early history or by
the long term enrichment by the group member galaxies that currently
occupy this region.  It is possible that the stars in NGC 5044 did not
produce iron with the same efficiency as in clusters, or the iron
resides in non-luminous clouds or stars, or the entropy of the
iron-enriched gas created in early galactic starburst winds was too
high to penetrate the group gas of lower entropy.
\end{abstract}


\section{Introduction}
\label{intro}

One of the most troublesome and least-discussed inconsistencies with
strict hierarchical cosmological scaling is the systematic difference
of metallicity between rich clusters and smaller galaxy groups from
which they presumably formed.  Most heavier elements are contained in
the hot intracluster gas and can be measured via X-ray
spectroscopy. Although these abundance differences may require
modification of IMF-averaged supernova yields, the sense of the
group-cluster discrepancy is that groups contain less metals per
$L_B$ than clusters \citep[e.g.,][]{renz97}. Similar conclusions were
found from gasdynamical models of groups
\cite[e.g.,][]{brig99a}.

The ratio of iron mass to optical light in rich clusters of galaxies,
which are expected to have retained all the metals produced by stars,
is $\Upsilon_{\rm Fe, gas} = M_{\rm Fe, gas}/L_B \sim 0.015$ in solar
units \citep[e.g.,][]{renz97,loew04a}.  Adding the iron locked in
stars, this increases to $\Upsilon_{Fe}\sim 0.02$. Here we assume an
average stellar mass to light ratio $\Upsilon_B = \langle M_*/L_B
\rangle =4$ and an average stellar iron abundance $\langle Z_{Fe}
\rangle = 0.7$ Z$_{Fe,\odot}$.  This value implies that about 0.0035
M$_\odot$ of iron are produced per M$_\odot$ of stars formed
\citep{kawa03a}. The ratio of iron mass to optical light
$(\Upsilon_{\rm Fe, gas})$ for groups is typically less than about
half that of rich clusters \citep[e.g.,][]{renz97}.
Even with the superior capabilities of \chandra\ and {\it XMM}, X-ray
observations of bright groups have yet to extend to the virial radius,
so uncertain extrapolations are required to estimate their total iron
and baryon content.

The strong radial iron abundance gradients observed in the hot gas of
many isolated galaxy groups are particularly interesting since they
suggest that this gas has been relatively undisturbed and unmixed by
merging.  Gas in the central parts of groups is enriched primarily by
Type Ia supernovae (SNIa) in central E or cD galaxies. However, more
distant gas is expected to be enriched by early galactic winds
from non-central, low luminosity group member galaxies. This raises
the interesting question whether or not the smaller amounts of iron
observed in hot gas near the periphery of groups can be understood as
enrichment by SNII and SNIa in non-central galaxies.

There is general agreement from both \rosat\ and \asca\ that
$\fe < 1\solar$ outside the cores of groups
\citep[e.g.,][]{fino99a,buot00c}. But because of limitations in
sensitivity and spectral or spatial resolution, these results were
considered tentative. Typically, \rosat\ and \asca\ found
$\fe/\solar=0.1-0.5$ out to $r\sim 100$~kpc. Recently, \citet{sun03a}
have reported iron abundances out to 200~kpc in the galaxy group NGC
1550.  Inspection of their figure 4 reveals (rescaled to the solar
abundances of \citealt{grsa}) that between 100-200~kpc
$\fe/\solar\approx 0.2$ but apparently with errors that allow for
values as high as $\fe/\solar= 0.3$.


Previously we reported abundances within the central region of the
X-ray luminous group NGC 5044 \citep[][ hereafter Paper 1 and
Paper 2 respectively]{buot03b,buot03a} finding $\fe/\solar \approx 1$
within $r\approx 30$~kpc and $\fe/\solar \approx 0.3$ near $r\approx
100$~kpc.  In this Letter we report accurate iron abundance
measurements out to $r\approx 330$~kpc ($\sim 0.4\rvir$) obtained from
a new \xmm\ observation of NGC 5044 offset from the center of the
group. We demonstrate that the very sub-solar values of
\fe\ we obtain at large radius are in serious conflict with 
metal enrichment models for galaxy groups, even in the presence of
group winds that eject baryons and heavy elements from the system.
Full details of the observations, enrichment models, and derived mass
distributions of stars, gas, and dark matter will appear in
forthcoming papers (Papers 3 and 4).  We assume a distance of 33 Mpc
to NGC 5044 for which 1$'$ = 9.6 kpc.

\section{Iron Abundance from XMM}
\label{iron}

\begin{figure*}[t]
\parbox{0.49\textwidth}{
\centerline{\psfig{figure=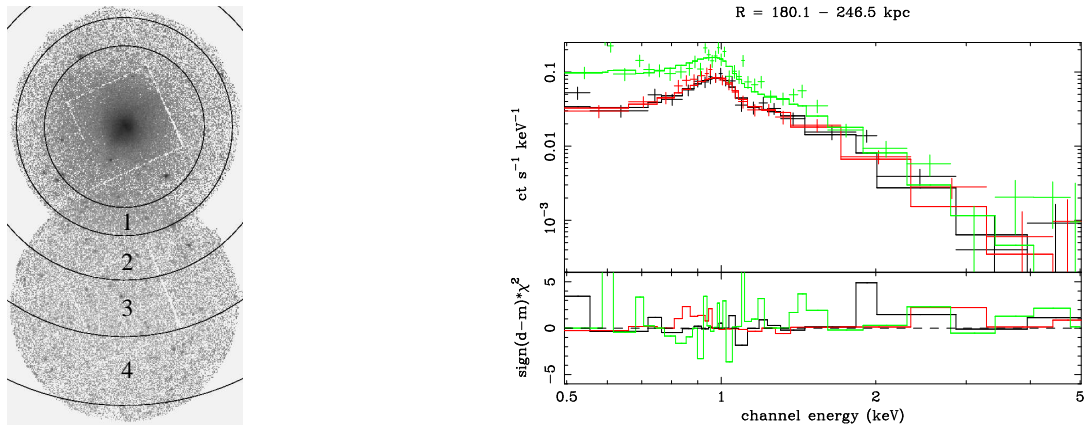,angle=0,height=0.23\textheight}}}
\parbox{0.49\textwidth}{
\centerline{\psfig{figure=f1b.eps,angle=-90,height=0.23\textheight}}}
\caption{\label{fig.data} \footnotesize
({\it Left}) EPIC MOS images (0.5-5~keV) for the AO-1 (top) and AO-2
(bottom) observations of NGC 5044. Overlaid are the annuli used to
extract the spectra discussed in this paper. ({\it Right}) The
0.5-5~keV spectra of the MOS (lower curve -- MOS1:black, MOS2:red )
and pn (upper curve, green) data in bin \#3 fitted with a single
\apec\ thermal plasma.}
\vspace{-0.5cm}
\end{figure*}

\begin{deluxetable*}{cllll|ccc|ccc}
\tablecaption{Temperature and Iron Abundance of the Hot Gas \label{tab.spec}}
\tablehead{
& & & & & \multicolumn{3}{c}{2D} & \multicolumn{3}{c}{3D}\\ 
& \multicolumn{2}{c}{$r_{\rm in}$} & \multicolumn{2}{c}{$r_{\rm out}$} & \colhead{$T$}& \colhead{$\fe$} & & \colhead{$T$}& \colhead{$\fe$}\\
\colhead{Bin}  & \colhead{(arcmin)} & \colhead{(kpc)} & \colhead{(arcmin)} & \colhead{(kpc)} & \colhead{(keV)} & \colhead{(solar)} & \colhead{$(\chi^2$/dof)} & \colhead{(keV)} & \colhead{(solar)} & \colhead{$(\chi^2$/dof)}
}
\startdata
1 & 10.0 & 94.8  & 13.5  & 128.0 & $1.06\pm 0.01$ & $0.22\pm 0.01$  & $1132.7/1008$  & $1.06\pm 0.01$ & $0.27\pm 0.02$  & $1125.5/1008$  \\
2 & 13.5 & 128.0 & 19.0  & 180.1 & $1.07\pm 0.02$ & $0.12\pm 0.01$  & $554.4/537$    & $1.22\pm 0.06$ & $0.13\pm 0.03$  & $552.3/537$    \\
3 & 19.0 & 180.1 & 26.0  & 246.5 & $0.98\pm 0.03$ & $0.16\pm 0.03$  & $727.5/678$    & $0.98\pm 0.03$ & $0.16\pm 0.03$  & $727.5/678$    \\
4 & 26.0 & 246.5 & 34.5  & 327.1 & $0.81\pm 0.03$ & $0.15\pm 0.05$  & $742.9/659$    & $\cdots$       & $\cdots$        & $\cdots$     
\enddata
\tablecomments{Results of fitting a single APEC plasma model (with
Galactic absorption) to the \xmm\ spectra. Quoted errors are
$1\sigma$. ``2D'' indicates no deprojection was performed. Radial bin
4 is not included in the deprojection analysis (see text).}
\end{deluxetable*}

The $22.5\arcmin$ observation offset due south of NGC 5044
(Figure \ref{fig.data}) was performed with the EPIC MOS and PN CCD
cameras for a nominal 40~ks exposure during AO-2 as part of the
\xmm\ Guest Observer program. The background was very quiescent for
the entire observation allowing for cleaned exposures near the nominal
value of 40~ks for all CCDs. Although no bright, extended sources
appear in the offset field, $\sim 50$ faint point sources were
detected and masked out of the ensuing analysis. 

Since background emission is a sizable fraction of the total flux in
the offset region, we modeled the background rather than using results
from blank fields. We followed previous studies of the X-ray
background with \xmm\ \citep[e.g.,][]{lumb02a} and modeled the
instrumental background using out-of field-of-view events for each
CCD. The cosmic X-ray background (CXB) was modeled using two thermal
plasma components for the hot Galactic halo and a power law for
unresolved point sources. Our CXB constraints are quite similar to
those reported in previous studies
\citep[e.g.,][]{lumb02a}.

We extracted spectra from each \xmm\ detector (MOS1, MOS2, PN) with
sections of circular annuli as shown in Figure \ref{fig.data} and
listed in Table \ref{tab.spec}. The center of these annuli coincides 
with that used Papers 1 and 2. (Note that the portions of the annuli
that do not lie on a detector are excluded from the analysis.)  Radial
bin 1 lies immediately outside the region studied in Papers 1 and 2
but is completely contained within the field of the AO-1
pointing. Consequently, we include the AO-1 data for radial bin 1 in
the present analysis and model the background similarly to that
described above for the AO-2 observation. Further details of the data
reduction and background analysis are discussed in Paper 3.

The source+background models were fitted simultaneously to the spectra
of each \xmm\ detector. The source models are APEC thermal plasmas
with temperature, iron abundance (\fe), and normalization as free
parameters. We use the new standard solar abundances of \citet{grsa}.
The source parameters and the normalizations of the background
components were fitted separately in each radial bin.  Results using
the best-fitting background model are quoted in this Letter.

The parameters of the source models obtained from fits to the data
without deprojection (i.e., ``2D'' models) are listed in Table
\ref{tab.spec}. A single temperature component provides a good
description of the data.  (The fit to the pn is not quite as good as
to the MOS due to imperfections in the background model. But excluding
key residuals, such as energies between 0.55-0.65~keV and above 3~keV,
gives an acceptable fit and provides temperature and \fe\ consistent
with the full-band analysis.) Even at these large radii the Fe L shell
lines near 1~keV are clearly visible (Figure \ref{fig.data}) and allow
for precise constraints on
\fe. Over most of the AO-2 field $\fe \approx 0.15\solar$ which is
much less than the (emission-weighted) average value $\fe=0.44 \pm
0.02\solar$ obtained over 48-96 kpc with the AO-1 data (Paper 2). We
note that these low \fe\ values are robust to errors in the background
model; e.g., for radial bin 3, we find that the $1\sigma$ error on
\fe\ increases from $0.03\solar$ to $0.05\solar$ when accounting for
statistical error in the background model parameters. Also, accounting
for weak Fe, O, and Ne lines due to solar
wind charge exchange emission \citep{snow04a} would reduce
slightly our measured \fe\ values, particularly in bin 4

Since \fe\ can be underestimated according to the ``Fe bias''
\citep[e.g.,][]{buot00a}, we investigated whether multi-temperature
spectral models and spectral deprojection made an important difference
in the fitted \fe\ values. Multi-temperature models (e.g.,
two-temperature) do not improve the fits significantly and do not
yield larger \fe\ values. Deprojecting the data (i.e., ``3D'' models
-- see Paper 1 for details) under the assumption of spherical symmetry
also does not significantly improve the fits (see Table
\ref{tab.spec}), though the value of \fe\ is raised by $0.05\solar$ in
radial bin 1. The assumption of spherical symmetry allows us to
extrapolate the spectra beyond the areas of the CCDs shown in Figure
\ref{fig.data} to the entire azimuthal range. Such extrapolation
magnifies azimuthal fluctuations in the source and background
intensity within the sections represented by the AO-2 field and
becomes increasingly more important in the outer radial bins: the area
extrapolation factors are $5-6$ for bins 1-3 and $\approx 9$ for bin
4. When bin 4 is included in the deprojection, the azimuthally
averaged gas density does not decrease monotonically with increasing
radius.  Consequently, we exclude bin 4 from the deprojection analysis
in this Letter.

These measurements currently represent the most accurate constraints
on \fe\ at radii 200-300~kpc from the center of any $T\sim 1$~keV
galaxy group and demonstrate $\fe\approx 0.15\solar$ at these large
radii. Below we use these measurements to constrain the global
chemical enrichment of NGC 5044.

\section{Inventory of Mass Components}
\label{mass}

\begin{deluxetable}{ccccc}
\tablecaption{Integrated Masses \label{tab.mass}}
\tablehead{
\colhead{$r$} & \colhead{\mgas} & \colhead{\mstars} & \colhead{$M_{grav}$} & \colhead{$M_{\rm Fe, gas}$} \\
\colhead{(kpc)} & \colhead{($10^{11}\msun$)} &
\colhead{($10^{11}\msun$)} & \colhead{($10^{13}\msun$)} &
\colhead{($10^{8}\msun$)}
}
\startdata
327 & $12.6\pm 0.08$ & $5.8$ & $2.00\pm 0.09$ & $3.5\pm 0.2$ \\
870 & 45.5 & 7.4 & 3.9 & 9.0
\enddata
\tablecomments{Errors are $1\sigma$. Quantities listed
within $\rvir=870$~kpc are extrapolations using data
within $r=327$~kpc.}
\end{deluxetable}

\begin{figure}[t]
\vskip -0.5cm
\centerline{\psfig{figure=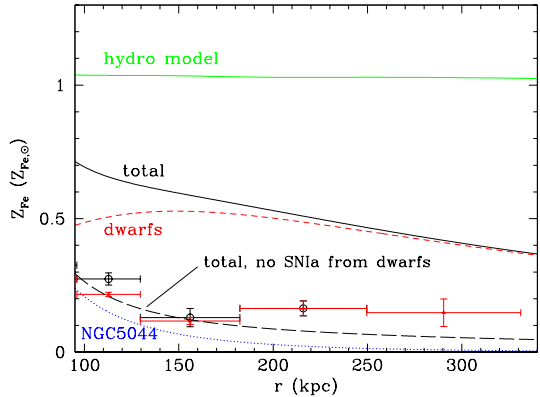,angle=-90,height=0.25\textheight}}
\vspace{-0.3cm}
\caption{\label{fig.models} \footnotesize
Gas iron abundances in NGC 5044.  New \xmm\ observations for $95\leq
r\leq 330$ kpc are shown with (open circles) and without (triangles)
deprojection.  The estimated on-the-spot total abundance (solid line),
is the sum of the gas enrichment due to the central E galaxy (dotted
line) and the ensemble of non-central member galaxies (short dashed
line). The upper solid line shows \fe\ of a typical gasdynamical model
with a supernova iron enrichment scenario similar to that of rich
clusters. The long-dashed line shows the on-the-spot abundance when
the gas is not enriched by SNIa in the non-central galaxies.}
\vspace{-0.5cm}
\end{figure}

Estimates of the expected hot gas iron abundance 
require an inventory of the various mass
components: hot gas, stars, dark matter and iron in the hot gas. 
The integrated masses listed in Table \ref{tab.mass} refer to
single-phase models of the hot gas except for $M_{\rm Fe, gas}$. Since 
two-temperature models are superior fits to \xmm\ and \chandra\
data within $r\la 30$~kpc and yield larger \fe\ values, the values of
$M_{\rm Fe, gas}$ in Table \ref{tab.mass} reflect two-temperature models in the
core of the group and single-temperature models elsewhere. However,
since the gas mass is weighted heavily at large radius, the value of
$M_{\rm Fe, gas}$ obtained using a single-temperature 
throughout is quite similar:
$\fem = 3.2\pm 0.2 \times 10^{8}\msun$.

We determined the gravitating mass assuming hydrostatic equilibrium
closely following the procedures outlined in
\citet{lewi03a} and \citet{buot04a}. The gravitating mass is well described by an
NFW \citep{nfw} model with concentration parameter, $c=11.1\pm 0.4$,
and virial radius, $\rvir= 870 \pm 0.02$~kpc. (Note: here \rvir\ is
defined within an overdensity of 103 appropriate for the $\Lambda$CDM
concordance model.) The value for concentration is within reasonable
scatter expected for CDM halos \citep[e.g.,][]{bull01a}.

To estimate the iron produced by supernovae, we require an estimate of
the radial profiles of stellar mass and density in the NGC 5044
group. The central elliptical galaxy can be fit with a de Vaucouleurs
profile \citep{goud94} with effective radius $R_e = 62.81$$''$ = 10.05
kpc. With RGC3 magnitudes and extinction $A_B = 0.30$ \citep{tonr01}
we find $L_B/L_{B,\odot} = 4.50
\times 10^{10}$.  This corresponds to a total stellar mass $M_{*,E} =
\Upsilon_{B,E} L_B = 3.4 \times 10^{11}$ $M_{\odot}$ where
$\Upsilon_{B,E} = M_{*}/L_B = 7.5$ is appropriate for a galaxy of this
$L_B$ \citep{gerh01a} using $H_0 = 70$ km sec$^{-1}$ Mpc$^{-1}$.

The central E galaxy is surrounded by $\sim 160$ smaller galaxies
($-19 \leq M_B \leq -13$) that we shall collectively refer to as
``dwarfs''. Only a few dwarf galaxies have known radial velocities.
Of the 163 group member galaxies tabulated by \citet[][hereafter
FS]{ferg90a} we have eliminated three (FS 68, 102 \& 137) because
their velocities differ by more than $\pm 800$ km s$^{-1}$ from NGC
5044. Luminosities $L_B$ have been determined from the magnitudes from
FS (with $A_B = 0.30$).  According to FS, most dwarf members are early
types (dE, dSph, S0, etc.) for which we assume $\langle B - V \rangle
\approx 0.7$ \citep{kari03a}.  Luminosities are converted to mass
with $\Upsilon_{B,dE} \approx 4.5$, typical for early type dwarfs
\citep{geha02a} and $\Upsilon_{B,dI} \approx 2.3$ for late types. The
total luminosity and mass of all non-central galaxies are $L_{B,d}
\approx 1.0 \times 10^{11}$ $L_{B,\odot}$ and $M_{*,d}
\approx 3.9 \times 10^{11}$ $M_{\odot}$.  The radial mass distribution
of the dwarf galaxies is fit with a King profile $\rho_{*,d} = \rho_0
[ 1 + (r/r_c)^2]^{-3/2}$ extending out to $r_t = 570$ kpc with $\rho_0
= 1.17 \times 10^4$ $M_{\odot}$ kpc$^{-3}$ and $r_c = 130$ kpc.
Finally, using parameterized fits to the integrated masses within
327~kpc we extrapolated the masses of each component to $\rvir \approx
870 $~kpc and list the results in Table \ref{tab.mass}. For $M_{\rm
Fe, gas}$ we used a power-law extrapolation based on values in radial
bins 1-3.

\section{Baryon and Iron Containment}
\label{contain}

From Table \ref{tab.mass} the total baryonic mass within the maximum
radius observed, $r_{obs} \approx 327$ kpc, is $M_{b}(r_{obs}) =
M_{gas}(r_{obs}) + M_{*}(r_{obs})
\approx 1.8 \times 10^{12}$ $M_{\odot}$.
The ratio of baryonic to total mass at $r_{obs}$ is therefore
$f_b(r_{obs}) \approx 0.09$.  The baryon fraction $f_b(\rvir) \approx
0.14$ extrapolated to the virial radius is
less certain but is about 85 percent of the cosmic value
$\Omega_b/\Omega_m = 0.044/0.27 = 0.16$.  Only about 15 percent of the
baryons have been lost from the NGC 5044 group, possibly by group
winds at early times if the SNII heating efficiency is sufficiently
high.

The total iron mass deficiency of NGC 5044 is much larger. The total
amount of stellar iron in the central galaxy is $M_{Fe,*,E} \approx
\langle Z_{Fe,E} \rangle (M_{*,E}/1.4) \approx 2.9 \times 10^8 $
$M_{\odot}$ where we assume $\langle Z_{Fe,E} \rangle \approx 0.7
Z_{Fe,\odot}$.  The total iron mass in the dwarfs, 
$M_{Fe,*,d}$, is only
1/2 of this, assuming $\langle Z_{Fe,d} \rangle = 0.24$ $Z_{Fe,\odot}$
\citep[e.g.,][]{rako01a}.
With these assumptions, 
the total mass of iron in both stars and gas within
$r_{obs}$ and $\rvir$ is $M_{Fe}(r_{obs}) \approx 6.5 \times 10^8$
$M_{\odot}$ and $M_{Fe}(\rvir) \approx 13.3 \times 10^8$ $M_{\odot}$.

If iron was initially produced in NGC 5044 with the same efficiency as
in rich clusters, we expect that the total mass of iron from all past
SNII and SNIa within $\rvir$ to be $M_{Fe,exp}
\approx 0.0035 M_{*}(\rvir) \approx 2.6 \times 10^9$ $M_{\odot}$, 
which exceeds the observed iron mass by at least a factor of $\sim
2$. The total iron mass to light ratio, $\Upsilon_{Fe}
\approx 0.009$, is about 2 times less than observed in rich clusters.
Evidently, about 50 percent of the total iron is unaccounted for in
NGC 5044. The amount of missing iron is much greater than the global
depletion of baryons, a result that is not very sensitive to our
extrapolation to $\rvir$.


Figure \ref{fig.models} illustrates the hot gas iron abundance
measurements for NGC 5044 within the 100 - 300 kpc
annulus of the offset region.
The iron abundance, $Z_{Fe}/Z_{\odot} \approx 0.1 - 0.2$, is
significantly lower than the typical abundance in rich clusters,
$Z_{Fe}/Z_{\odot} \sim 0.4$.
Is the low $Z_{Fe}$ in this distant gas in NGC 5044 consistent with
iron-enriched outflows from the dwarf galaxies in this same region?
For an approximate answer to this question we assume that the local
gas was enriched by dwarf galaxy winds according to an ``on-the-spot''
scenario, where the dwarf galaxies (and distant stars of the central E
galaxy) are surrounded by the same ambient gas that accompanied them
when they first entered the group halo.  Orbital models for dwarf
member galaxies show that their time-averaged radial positions lie
near the virial radius at the time when they first joined the group.
In the on-the-spot approximation, the iron
abundance in the hot gas is found by subtracting the iron in stars
from the total iron produced by all supernovae over time,
$$
Z_{Fe,ots}(r) = 
{1 \over Z_{\odot}}~\sum_{i=E,d}
\left[ {0.0035 \over 1.4} - Z_{Fe,*,i}(r)\right]
{\rho_{*,i}(r) \over \rho_{gas}(r)} ,
$$
where the two terms represent the iron produced by the central E
galaxy ($i = E$) and the surrounding cloud of dwarf galaxies ($i =
d$).
The stellar iron abundance in the central E galaxy is approximated
with $Z_{Fe,*,E}(r)/Z_{\odot} \approx 0.675\,(r_{kpc}/10.05)^{-0.21}$
\citep{koba99}
although the central galaxy does not appreciably enrich the offset
region $100 < r < 300$ kpc.

The on-the-spot iron abundance $Z_{Fe,ots}(r)$ in the offset region
plotted in Figure 2 is seen to exceed the observed abundances by
factors of 3 - 4.  Indeed this discrepancy may be {\it
underestimated} since on-the-spot enrichment does not include iron
expelled from former massive group member galaxies (brighter than the
most luminous surviving non-central galaxy, $M_B = -19$) that were
originally in the offset region but have since merged with the central
E galaxy by dynamical friction. The iron abundance observed in $r >
100$ kpc is also very much less than that predicted by standard 1D
gasdynamical models of early group winds in which metals are uniformly
mixed with the ICM \citep{brig99a}.  
The iron enrichment expected in a typical gasdynamical model using
rich cluster SNII rates shown in Figure 2 results in very large
$Z_{\rm Fe, gas}$ beyond 100 kpc.  
This is due to (1) the outward
advection of supernova products from the massive central galaxy and
(2) the increased efficiency of increasing $Z_{\rm Fe, gas}$ by SNIa
enrichment when the local gas density is lowered by a global group
wind. Finally, accretion of (possibly iron-poor) primordial gas cannot
easily account for the low observed $Z_{Fe,gas}$. Cosmologically
accreted gas, which is included in our dynamical models, flows
differentially inward past the dwarf galaxies by less than $70$ kpc
over $10^{10}$ years.  Shocked gas currently arriving at $r_{vir}$
cannot mix deeper into the group IGM because of its higher entropy.

Both the on-the-spot approximation and the detailed gasdynamical
calculations over-predict the iron abundance in the offset region.  We
offer several possible explanations: (1) the stars in NGC 5044 did not
produce iron with the same efficiency as in clusters; in particular,
Type Ia supernovae in non-central galaxies do not occur at the
expected rate or failed to enrich the local intergalactic gas, (2)
iron has been selectively ejected from the group or resides in
non-luminous clouds or stars, or, (3) high entropy gas enriched and
heated by early SNII and SNIa explosions may not have penetrated
deeply inside the NGC 5044 group because of its inherent buoyancy.





\acknowledgements We thank the referee, F. Paerels, for helpful
comments, A.\ Lewis for assistance, and 
acknowledge support from NASA grants NAG5-13143 and NAG5-13059.


\end{document}